\documentclass{ws-ijmpc}

\usepackage{graphicx}

\begin{document}

\title{Continuous Opinions and Discrete Actions in Opinion Dynamics Problems}
\author{Andr\'e C. R. Martins}
\address{GRIFE - Escola de Artes, Ci\^encias e Humanidades\\
Universidade de S\~ao Paulo, Brazil\\
amartins@usp.br}

\maketitle

\begin{abstract}
A model where agents show discrete behavior regarding their actions, but have continuous opinions that are updated by interacting with other agents is presented. This new updating rule is applied to both the voter and Sznajd models for interaction between neighbors, and its consequences are discussed. The appearance of extremists is naturally observed and it seems to be a characteristic of this model.

\end{abstract}



\section{Introduction}

Opinion dynamics models~\cite{odreview} can be divided into two groups. On one side, there are models where the opinions are considered discrete (often accepting only two different results). In this group, per example, we have the first uses of the Ising model to describe the behavior of laborers in a strike~\cite{galam82} and the emergence of consensus~\cite{galam91}; the voter model, where each agent is influenced by one of its neighbors at a time~\cite{voter1,voter2}; and the Sznajd model, where it takes two agents to convince their neighbors of the correctness of their opinion~\cite{sznajd2000,staufer2003,sznajd2005}. On the second group, that of continuous opinions, we have, per example the Deffuant~\cite{deffuant} and the Hegselmann-Krause~\cite{hegselmann} models. Both models used the concept of bounded confidence, that is, agents only influence each other if the difference in their opinions is not larger than some threshold and the models differ in how many other agents each agent can interact with at a certain time. 

While the discrete models are useful to represent situations where binary choices are a good description of the problem, the agents have no memory of their past opinions. One proposal to deal with this problem was the use of active Brownian particles~\cite{schweitzer}, particles that can store energy in an internal depot and use this energy to influence their future behavior. Another problem with discrete models is that they are not well suited to describe the emergence of extremism in the system~\cite{deffuant2002}, since opinions have only two values and, therefore, no extremal opinions.

In this article, a model for opinions that are observed as discrete actions but are represented internally by each agent as a continuous opinion function is proposed. This function can be understood as associated with the probability that the agents assign to the statement that one of the two available alternatives is the best one. A simple updating rule can be obtained by assuming that the agents change their opinions by using a very simple Bayesian description of how likely their neighbors are to be correct. This update rule is implemented and the consequences of this rule for the voter and the Sznajd model are studied in regular square lattices. Section~\ref{sec:coda} defines the Continuous Opinions and Discrete Actions (CODA) model, the model is applied to the voter model in Section~\ref{sec:voter} and to the Sznajd model in Section~\ref{sec:sznajd}.

\section{Continuous Opinions and Discrete Actions}\label{sec:coda}

When someone faces a binary decision, the opinion about which option is the best one is not necessarily binary. For most problems, it is reasonable to assume that the person believes one of the alternatives is better with a probability $p$. If the consequences of being right or wrong are equivalent for both choices, the alternative with higher probability, $p$ or $1-p$, will be chosen as the best one. Under circumstances where each individual notices only the choices of other individuals, but is not aware of their internal opinions, there is no way that interacting agents will converge to a mean result as in the bounded confidence models~\cite{deffuant,hegselmann}. Instead, each agent will change its continuous internal probability towards the value of its peers. Therefore, some people will change their actions after one interaction, while others, with more extreme opinions, might take several interactions with people with opposed actions before they actually change their own public choices.

Of course, a simple additive random walk over the probability value will not work, since probability must be limited between zero and one. Luckily, a simple application of Bayes theorem suggests an easy way to do the updating. Assume that the probability in favor of the first alternative, $A$, is $p_i$ and, therefore, the probability favoring the second alternative, $B$, is $1-p_i$. From now on, the index $i$ will not be used, for simplicity of notation. If an agent observes someone who acts as if she believed in $A$, this should change $p$ to a new larger value. The choices will be represented as a discrete field $\sigma_i(p)$, that accepts two values, $\sigma_i=+1$, if the agent $i$ chooses $A$ and $\sigma_i=-1$ if she prefers $B$.

 In order to implement a Bayesian update, we need the likelihoods associated with the actions of the neighbors. Therefore,  $\alpha=P(\sigma_j=+1|A)$ is the probability that neighbor $j$ chooses $A$, if $A$ is true. Likewise, if $B$ is the best alternative, there is a $\beta=P(\sigma_j=-1|B)$ probability that the agent would choose also $B$ (or a $1-\beta$ probability the neighbor will chose $A$).  Given the prior probability $p$ and the likelihoods, the probability $P(A|\sigma_j=+1)$ that $A$ is the best alternative after observing someone that supports $A$, will become $P(A|\sigma_j=+1)\propto P(A)P(\sigma_j=+1|A)=p\alpha$. 

In order to get rid of the normalizing constant, it is easier to deal with the odds in favor of $A$, $O(A)$. The odds is defined as the ratio between the belief in $A$ and the belief in $B$, cancelling normalizing constants. The prior odds $O(A)$, when $P(A)=p$ will be given by
\begin{equation}
O(A)=\frac{P(A)}{P(B)}=\frac{p}{1-p}.
\end{equation}
For the posterior odds $O(A|\sigma_j=+1)$, after observing a neighbor that supports $A$, we have
\begin{equation}\label{eq:bayes}
O(A|\sigma_j=+1)=\frac{P(A|\sigma_j=+1)}{P(B|\sigma_j=+1)}=\frac{p}{1-p}\frac{\alpha}{1-\beta}.
\end{equation}

Notice that the first term is the prior odds, that is, after each observation, the initial odds opinion should be multiplied by $\frac{\alpha}{1-\beta}$. Here, we can make the reasonable assumption that no option $A$ or $B$ is somehow favored, that is $\alpha=\beta$. Notice that it is easier to deal with the logarithm of $O(A)$, a quantity known simply as log-odds. The changes in log-odds $l=\ln(O(A))$ will be additive, changing, at each step, by $\nu=\ln\frac{\alpha}{1-\alpha}$ and we have
\begin{equation}\label{eq:logodd}
l(A|\sigma_j=+1)=l(A)+\nu.
\end{equation}
It is important to notice at this point that $l$ is an invertible function of $p$ and, therefore, just another way to measure the probability of $A$. It will be used here simply because it is updated in a much simpler way than $p$, under Bayesian update rules.

As we will see later, the absolute value of $\nu$ will only be relevant in comparison with the initial opinions. After the system has reached an almost stable state in terms of actions (as we will see, opinions may diverge indefinitely), only the sign of $\nu$ will be relevant. Since it is reasonable to assume that $\alpha>0.5$, this leads to $\nu>0$. This means that whenever one observes someone that prefers $A$, $\nu$ and therefore $p$ becomes larger, while $p$ decreases when someone is observed who chooses $B$. The case where $\alpha<0.5$ can be interpreted as the belief that other people are more likely to pick the worst alternative than the best one and will make an agent a contrarian~\cite{galam2004}. Since $p=0.5$ translates to $l=0$, we can work just as easily with $l$ as with $p$ and the binary discrete actions will correspond to negative and positive values of $l$. If $l$ is exactly zero, one action must be chosen and, therefore, favored. However, since $l$ is continuous, this choice corresponds to a null measure and should not make any significant difference.

One should notice that, from the point of view of the agents, what they observe is only the actions. Those actions can be described as a field of spins $\sigma_i$ with exactly the same structure as the Ising model. That is true from the point of view of the agents. However, the model has another layer behind those actions and that is the continuous opinions of each agent, represented by the log-odds $l$. For the purpose of analysing the appearance of extreme opinions, it is the value of $l$ that must be measured. Since $l$ is the transformed probability $p$, it is also $l$ that matters for the real people the model is supposed to simulate. This happens because if someone must make a real decisions that is more complex than just telling which option she thinks it is best, that decision will depend on the exact value of $p$, not only on the sign of $l$. Therefore, the model gives more than just an Ising model with a different rule for updating the spins.

\section{Voter Model}\label{sec:voter}

The dynamics for opinion change described in the previous section can easily be implemented using the different models for interaction available. Per example, in the voter model~\cite{voter1,voter2}, an agent $i$ is randomly drawn at each step, together with one of its neighbors, $n_i$. The agent is influenced by the neighbor, adopting the neighbor opinion. For the CODA model, the neighbors still influence the agent, but by changing its internal probability associated with the action by the amount $\nu$. That way, if the agent and the neighbor disagree on their actions, the agent opinion will become less extreme and the agent might change its mind, if $l(i)$ was close enough to 0. On the other hand, if both agree, the agent will validate its choice and change its log-odds opinion to further away from 0. Eventually, this may turn some agents into extremist, who have very strong opinions ($p$ very close to 1 or 0, meaning a very large absolute value for $l$).

\begin{figure}[bbb]
 \includegraphics[width=12.5cm]{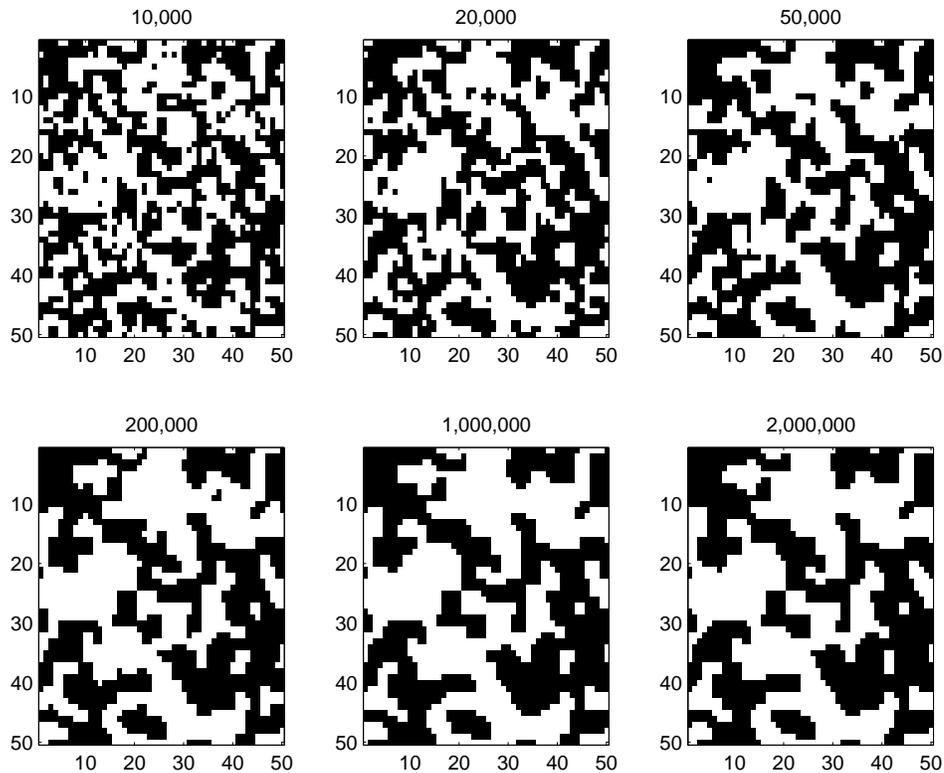}
  \caption{The evolution of the opinions of agents on a $50\times 50$ square lattice for voter interactions. The number of interactions shown for each configuration is the number of times one single agent was drawn and updated.}
\label{fig:voteropinion}
\end{figure}

Figure~\ref{fig:voteropinion} shows the evolution of configurations of a typical run with periodic boundary conditions after different numbers of agent updatings. Initial conditions were chosen so that $p$ was randomly drawn for each agent from two uniform distributions, between $0.4\leq p \leq 0.49$ and $0.51\leq p\leq 0.6$ and the update step was chosen as $\alpha=0.7$, meaning a change in the log-odds of $\nu=0.8473$ towards the direction favored by the neighbor in each interaction. The initial opinions ensure that, in average, an equal number of agents should favor each of the actions and they are distributed in a way that is not too far from flipping to the opposite direction. Also, notice that  the most extreme initial opinions $p=0.4$ correspond to $l=-0.4055$ (similarly for $p=0.6$, with the opposite sign) and, since the value is a little less than half the size of the opinion change, one single step can change the most radical possible initial opinion to its opposite. That is, at the beginning, there are no extremists.

\begin{figure}[bbb]
 \includegraphics[width=10.0cm]{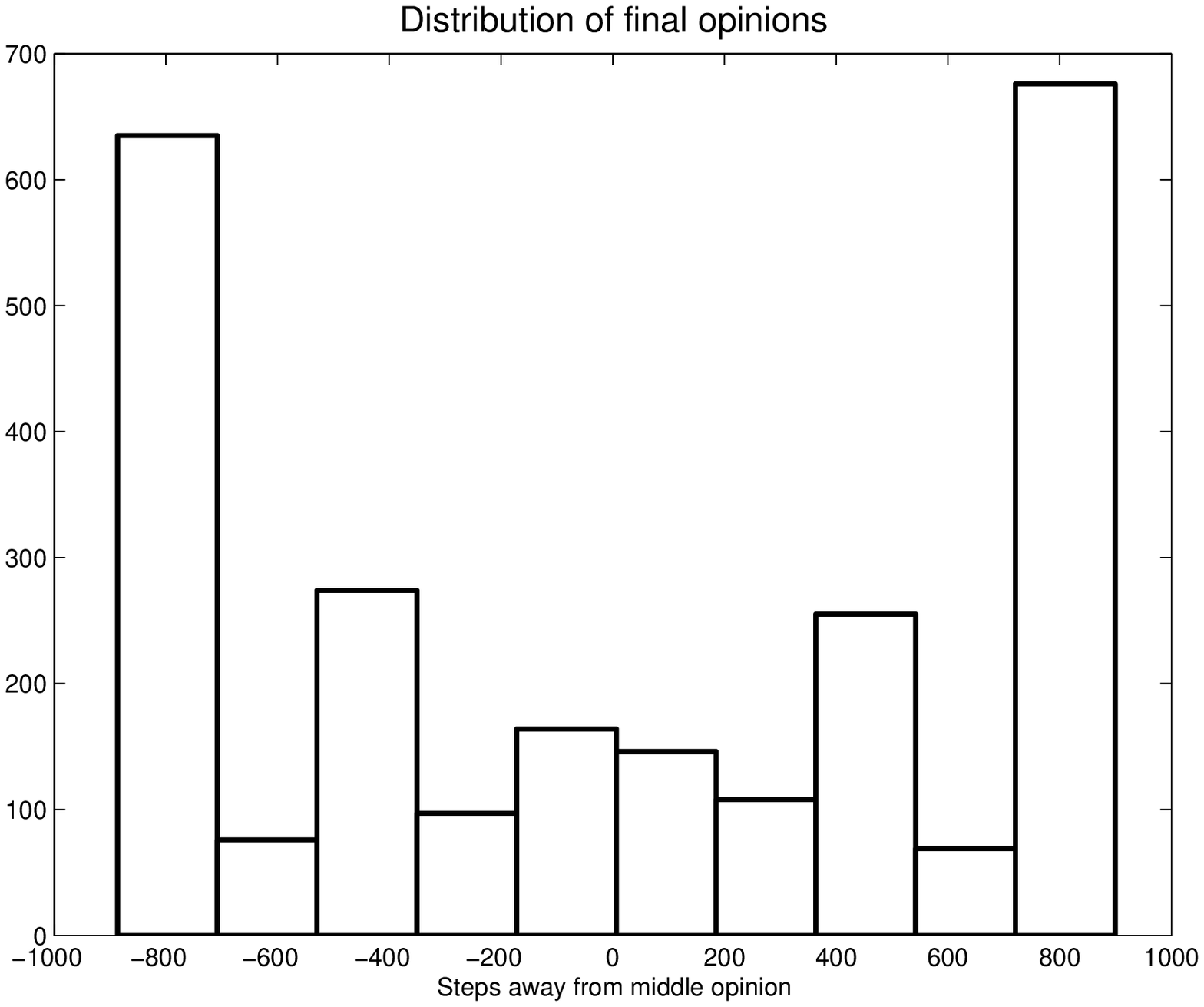}
  \caption{Distribution of the opinions after 2,000,000 of updates for voter interactions, measure as multiples of $\nu$, that is how many update steps each agent is from flipping opinion.}
\label{fig:voterdistribution}
\end{figure}
 
Since it takes 2,500 updatings to get all agents to change, the configurations correspond to what is observed after updating each agent 4, 8, 20, 80, 400 and 800 times on average. Notice that reasonably stable domains are formed, but small changes can be observed even after those domains are well established. 

As this is a simple model, the evolution of shape of the domains can be easily described. Straight lines between different domains are reasonably stable, since each agent has exactly 3 neighbors that agree with him and one who doesn't. This mean that it will interact with someone with the same opinion $3/4$ of the time, for an average change in $l$ of $\frac{3}{4}\nu-\frac{1}{4}\nu=\frac{\nu}{2}$. The standard deviation of one interaction is $\frac{\sqrt{3}}{2}\nu$ and, since it scales with $\sqrt{t}$, where $t$ is the number of times the agent updates its opinion, the wall will tend to get stronger in the long run, with both sides becoming more and more extreme in their opinions. This is not observed for corners, since, on average, the opinion change is zero.

Figure~\ref{fig:voterdistribution} shows the distribution of opinions of the agents after 2,000,000 individual agent updatings. The opinion is measured in the logarithm of the odds and as a multiple of $\nu$, that is it represents how many opinion changes in the right direction would be necessary in order to change the action choice to its opposite. Notice that real extremists, with a distance to the opposite opinion of around 800 steps are quite common at both sides. In order to understand how extreme that position is, for the step size of $\nu=0.8473$, an opinion 800 steps away from flipping corresponds to a probability of $p=4.2\cdot 10^{-295}$. Such a number, for any practical purpose, is equivalent to zero. Such an unreal certainty is associated with the value of $\alpha=0.7$. If the likelihood associated with the probability that the neighbor will choose the right option drops to $0.51$, 800 steps away from the flipping opinion translate to $p=1.26\cdot 10^{-14}$, still a very extreme opinion and it becomes just a little less than $4\%$ if $\alpha=0.501$. However, regardless of the value of $p$, it is still 800 steps away from changing opinions.

\section{Sznajd Model}\label{sec:sznajd}

\begin{figure}[bbb]
 \includegraphics[width=12.5cm]{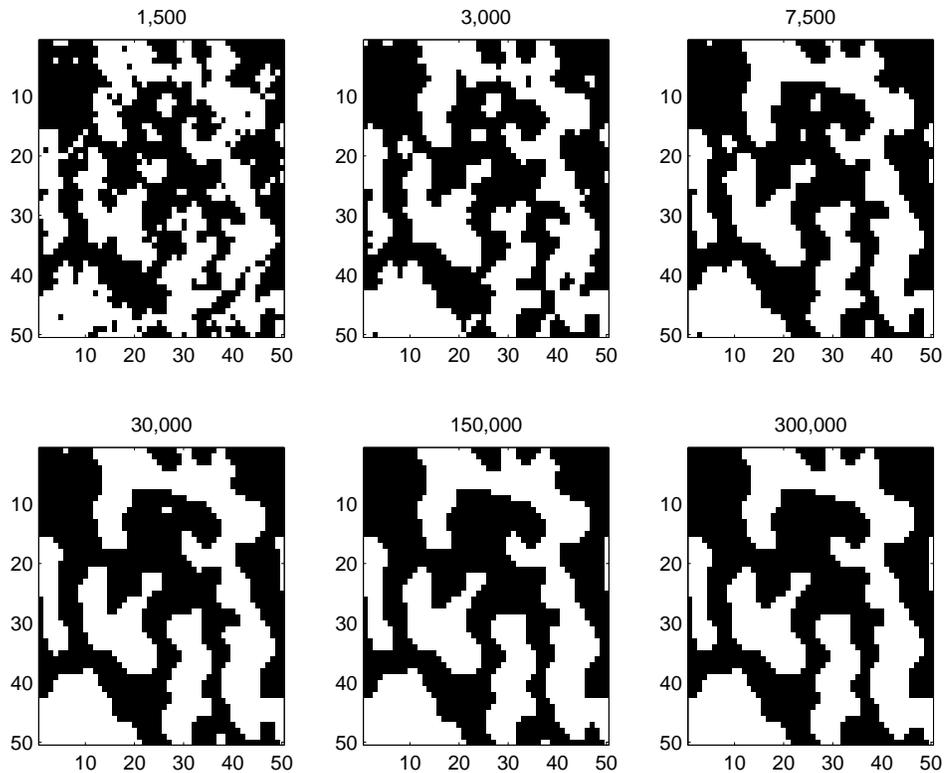}
  \caption{The evolution of the opinions of agents on a $50\times 50$ square lattice for Sznajd interactions. The number of interactions shown for each configuration is the number of times one single agent was drawn and updated.}
\label{fig:sznajdopinion}
\end{figure}

The CODA model was also applied to the Sznajd interaction rules~\cite{sznajd2000,staufer2003,sznajd2005}. Under those rules, the update is applied to more agents at each one time. The more common version of the Sznajd interactions imposes that when the two neighbors are randomly drawn, if they agree, they influence all their other neighbors. That is, for bi-dimensional lattices, six agents have their opinions updated at once. If they disagree, nothing is done. The results of the simulation for the Sznajd model with CODA updating rules were basically the same as those of the voter model, suggesting that the characteristics discussed in the Section~\ref{sec:voter} might be a general feature of the model.

\begin{figure}[bbb]
 \includegraphics[width=10.0cm]{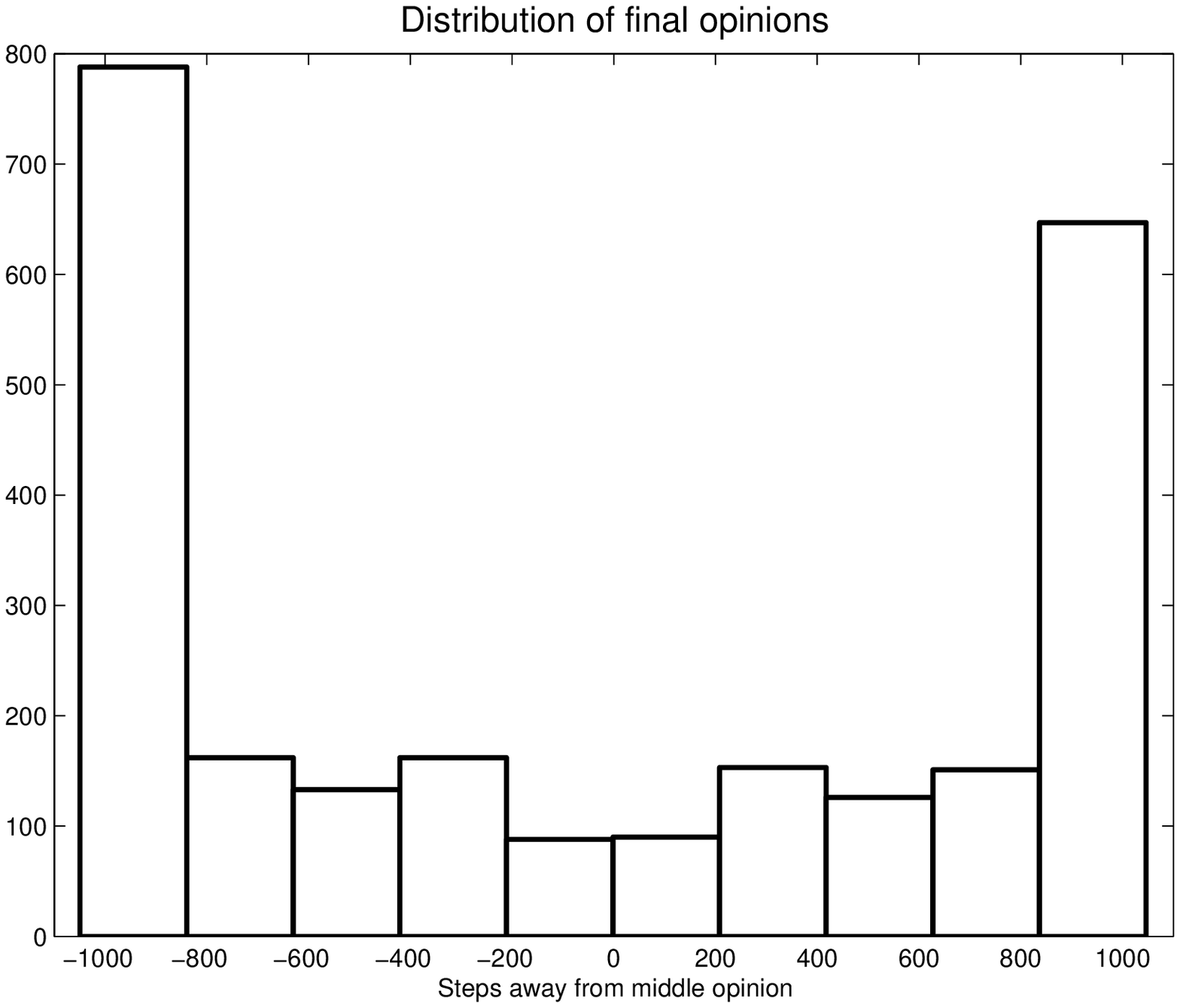}
  \caption{Distribution of the opinions after 300,000 of updates for Sznajd interactions, measure as multiples of $\nu$.}
\label{fig:sznajddistribution}
\end{figure}
 
The fact that the updating rule applies to a larger number of agents at each iteration allows for a much faster convergence and appearance of the domains, as can be seen in Figure~\ref{fig:sznajdopinion}. If the program is left to run for a longer time, some of the agents at the boundaries of the two opinions sometimes change their minds, as it had already happened in the voter model, but those changes happen very slowly.

The distribution of the opinions between extremists and centrists for the Sznajd rules can be seen in Figure~\ref{fig:sznajddistribution}. The behavior is basically the same of the voter model, with the very extreme opinions appearing as the most frequent ones. And if the agents are left to interact for longer, the inner regions, where all neighbors share the same opinion will again continuously reinforce themselves, driving the extreme opinions there to even more extreme values, repeating the behavior already obsreved for the voter model.

\section{Conclusion}

The CODA updating rule is proposed as an alternative to the usual updating rules, where opinions just flip, with no memory. By differentiating between opinion and choice, the rule allows to model easily the opinions of agents as continuous functions even when the observed actions are binary. Also, since this rule is obtained from a simple application of the Bayes rule, it is probably a good approximation for the opinions of rational agents under these circumstances.

We have seen that the application of this rule for agents located in square lattices with periodic boundary conditions to two different interaction models, the voter model and the Sznajd model, lead to very similar results. This suggests that the observed behavior might be a characteristic of the CODA rules that is preserved for different types of interactions between the agents.

Clear domains with different opinions were observed in both models, showing that the two opinions will have to live with each other for very long times. Also, inside those domains and, to a lesser extent, in the boundaries, opinions become very extreme, with each agent basically sure that his choice is the best one. This can help explain cases where people are led, by social pressure, to believe blindly in whatever opinion is shared by its local group, despite divergent voices in the larger society they live in.

\end{document}